# Reweighting of Binaural Localization Cues Induced by Lateralization Training

Maike Klingel (maiden name: Ferber)[1,2,3] (maike.ferber@univie.ac.at), Norbert Kopčo[3] (norbert.kopco@upjs.sk) and Bernhard Laback[1] (bernhard.laback@oeaw.ac.at)

[1]Acoustics Research Institute, Austrian Academy of Sciences, 1040 Vienna, Austria

[2]Department of Cognition, Emotion, and Methods in Psychology, Faculty of Psychology, University of Vienna, 1010 Vienna, Austria

[3]Institute of Computer Science, Faculty of Science, P. J. Šafárik University in Košice, 04180 Košice, Slovakia

**Corresponding authors:**

Maike Klingel

maike.ferber@univie.ac.at

Tel: +43 1 51581 2521

Fax: +43 1 51581 2530

Bernhard Laback

bernhard.laback@oeaw.ac.at

Tel: +43 1 51581 2514

Fax: +43 1 51581 2530

**Word count:**

Abstract: 266

Introduction: 986

Discussion: 2169

**Abstract**

Normal-hearing listeners adapt to alterations in sound localization cues. This adaptation can result from the establishment of a new spatial map of the altered cues or from a stronger relative weighting of unaltered compared to altered cues. Such reweighting has been shown for monaural vs. binaural cues. However, studies attempting to reweight the two binaural cues, interaural differences in time and level, yielded inconclusive results. In this study we investigated whether binaural cue reweighting can be induced by a lateralization training in a virtual audio-visual environment.

20 normal-hearing participants, divided into two groups, completed the experiment consisting of a seven-day lateralization training in a virtual audio-visual environment, preceded and followed by a test measuring the binaural cue weights. During testing, the participants' task was to lateralize 500-ms bandpass-filtered (2-4 kHz) noise bursts containing various combinations of spatially consistent and inconsistent ITDs and ILDs. During training, the task was extended by visual cues reinforcing ITDs in one group and ILDs in the other group as well as manipulating the azimuthal ranges of the two cues.

In both groups, the weight given to the reinforced cue increased significantly from pre- to posttest, suggesting that participants reweighted the binaural cues in the expected direction. This reweighting occurred predominantly within the first training session. The present results are relevant as binaural cue reweighting is, for example, likely to occur when normal-hearing listeners adapt to new acoustic environments. Similarly, binaural cue reweighting might be a factor underlying the low contribution of ITDs to sound localization of cochlear-implant listeners as they typically do not experience reliable ITD cues with their clinical devices.

**Introduction**

Spatial hearing is crucial for our ability to localize sound sources, to understand speech in complex environments and to orient in space. It is, therefore, a vital part of our everyday lives for which we use a combination of both monaural and binaural cues. For vertical-plane localization, we rely primarily on monaural spectral-shape cues, while for azimuthal sound localization, the topic of this study, the two binaural cues (interaural time difference, ITD, and interaural level difference, ILD) are most important (for a review on localization cues, see, e.g., Middlebrooks & Green, 1991).

The extent to which the two binaural cues contribute to azimuthal sound localization mainly depends on the frequency content of the sound. ITDs are dominant for low frequency sounds, up to approximately 1.5 kHz, whereas ILDs are dominant for higher frequencies (Macpherson & Middlebrooks, 2002; Strutt, 1907). The relative weight with which ITD and ILD contribute to a spatial percept has traditionally been measured using the ITD/ILD trading ratio approach, estimated by presenting one binaural cue at a fixed magnitude and instructing the participants to adjust the other cue until the sound is perceived centrally (Durlach & Colburn, 1978). However, trading ratios do not only depend on the frequency content, but also on other sound properties such as overall intensity (David, Guttman, & van Bergeijk, 1959; Deatherage & Hirsh, 1959) or the inter-click interval of click trains (Stecker, 2010). In addition to these stimulus factors, contextual or environmental factors influence the ITD/ILD weighting. Specifically, Rakerd and Hartmann (2010) observed that in reverberant environments, participants' responses follow ILDs while localizing 500 Hz sine tones, a stimulus for which localization is based on ITDs in anechoic environments. Moreover, ITD/ILD trading ratios depend on which cue is adjusted. Namely, this method leads to a greater weight of the to-be-adjusted cue, presumably because attention is shifted towards it (Lang & Buchner, 2008).

Furthermore, the relationship of sound localization cues to corresponding locations in space may change during one's life and, therefore, adaptation to these changes is necessary. Increasing head size during development (Clifton, Gwiazda, Bauer, Clarkson, & Held, 1988) or temporary physiological changes due to middle ear infections (Morrongiello, 1989), for example, affect sound localization cues reaching the cochleae. Consequently, adaptation to altered spatial cues has been extensively studied (see Carlile, 2014, and King et al., 2011, for reviews), highlighting the plasticity of the auditory system. Observed adaptation can either result from the establishment of a new spatial map of the altered cues (i.e., a modified relationship between sound localization cues and corresponding locations in space; Keating, Dahmen, & King, 2015; Knudsen, 2002; Shinn-Cunningham, Durlach, & Held, 1998; Trapeau & Schönwiesner, 2015) or from a stronger relative weighting of unaltered compared to altered spatial cues, referred to as reweighting (Keating, Dahmen, & King, 2013; Kumpik, Kacelnik, & King, 2010; van Wanrooij & van Opstal, 2007). In these latter studies, listeners learned to increase the relative weight for monaural compared to binaural cues for azimuthal sound localization. However, studies attempting to change the relative weight of the two binaural cues ITD and ILD yielded inconclusive results. An early study using a left/right discrimination task with spatially inconsistent ITD and ILD cues (one pointing to the left and the other to the right) in which feedback consistent with only one of the cues was presented did not show any changes in the ITD/ILD weighting after training (Jeffress & McFadden, 1971). While a recent study (Kumpik, Campbell, Schnupp, & King, 2019) observed an increase in ILD weighting after ITDs were randomized during a visual oddball task, it did not show an increase in ITD weighting after ILDs were randomized, and an increase in ILD weighting was also found for a condition in which no cue was randomized (i.e., spatially consistent ITDs and ILDs were presented), making it difficult to draw strong conclusions. Considering the dynamically changing relative contributions of ITD and ILD depending on sound properties and contextual factors as discussed above, these inconclusive results are rather surprising and worthy of further consideration. In Kumpik et al.'s (2019) study, the auditory stimuli were task irrelevant and therefore might not have received enough attention. The task in Jeffress and McFadden's (1971) study was a simple left/right discrimination, the stimuli were noise bands centered at 500 Hz, and the binaural cues were close to the binaural cue threshold. Therefore, potential reasons for their null result are that (1) the training regime was not sufficiently intuitive, (2)

REWEIGHTING OF BINAURAL CUESthe stimulus was in a frequency range where only ITDs but not ILDs arise naturally, and (3) binaural cues were not sufficiently salient.

Here, we reexamine the hypothesis of binaural cue reweighting by attempting to provide more favorable conditions for inducing an increase in either ILD or ITD weighting. We trained listeners using a well-controlled virtual environment involving auditory, visual and proprioceptive information based on the procedure used in Majdak, Walder and Laback (2013). The training uses two forms of visually-guided auditory spatial calibration: 1) visual stimuli presented after the auditory stimuli serving as top-down feedback comparable to the "feedback" experiments of Shinn-Cunningham et al. (1998), and 2) simultaneously presented auditory and visual stimuli to encourage multisensory bottom-up processes equivalent to those evoked in the ventriloquism aftereffect paradigm (Reccanzone, 1998). In contrast to Kumpik et al.'s (2019) study, the auditory stimuli were critical to perform the task and unlike Jeffress and McFadden (1971), we used a stimulus spectrally focused at an intermediate frequency region, providing salient ILD as well as ITD cues. Since the sensitivity to binaural cues can be modified based on the statistics of the sound (Dahmen, Keating, Nodal, Schulz, & King, 2010), we additionally manipulated the across-trial stability of the two binaural cues during training by varying one of the cues over a larger range than the other. Finally, a variety of combinations of ITDs and ILDs was presented to facilitate generalization of reweighting across spatial configurations and to minimize chances for strategic responding.

**Methods**

*Participants*

Twenty normal-hearing adult participants (10 female, mean age: 26.9 years, SD = 4.13, range: 21-40 years) completed the experiment. All participants gave informed consent before participating and received monetary compensation for their participation. Basic lateralization ability as well as sensitivity to both ITD and ILD were assessed in a practice session (see procedure for details) and served as an inclusion criterion. We used two experimental groups: ITDs were reinforced for the ITD target group and ILDs were reinforced for the ILD target group. Participants were pseudo-randomly assigned to one of the groups, ensuring balanced age, sex and basic lateralization ability. Ten participants (five female, mean age: 27 years, SD = 5.27), were assigned to the ITD target group. The other ten participants (five female, mean age: 26.8 years, SD = 2.86), were assigned to the ILD target group.

*Apparatus and Stimuli*

Participants stood on a platform surrounded by a circular railing inside a sound booth. The experiment was controlled by custom-written software routines that were run on two communicating personal computers (via UDP protocol), sharing the tasks of acquiring head tracking data, creating and presenting acoustic stimuli, and rendering the visual environment. Binaural auditory stimuli were generated using a computer and output via a digital audio interface (ADI-8, RME) at a 96-kHz sampling rate and presented via headphones (HD 580, Sennheiser). Visual stimuli were presented binocularly via a head-mounted display (Oculus Rift DK1). Participants' head position and orientation were tracked with a head-mounted tracking sensor (Flock of Birds, Ascension) and the visual environment was rendered accordingly in real-time (the latency between head movements and the updated visual information was less than 37.3 ms). The rendering of the visual environment using Pure Data (with GEM, IEM, Graz) was based on the left/right rotation information from the tracking sensor while the up/down information was ignored to force listeners to respond only in the horizontal plane. The virtual visual environment consisted of a reference position straight ahead, a crosshair in the direction of the head orientation, a single horizontal line at eye level, and vertical lines every 15° in azimuth for guidance (Figure 2). A rotating cube was used as visual reinforcement.

Auditory source stimuli were white noise bursts, randomly generated on each trial, which were filtered with an 2-4 kHz[1] butterworth band-pass filter (roll-off outside the passband: approximately 30 dB/oct; Fig. 1a), on which ITDs ranging from -662 to +662 µs and ILDs ranging from -19.4 to +19.4 dB were imposed. These cues correspond to an azimuthal range from -70.2° to +70.2° (Fig. 1b), as determined by Xie (2013) using the head related transfer functions (HRTFs) of the KEMAR head with DB-61 small pinna at a source distance of 1.4 m. In Xie (2013), the ITD values were obtained via broadband cross-correlation of the left and right ear head-related impulse responses (HRIRs) and the ILD values based on the HRTF magnitudes at 2.8 kHz (i.e., the center frequency of the auditory stimuli used in the present study). The choice of the stimulus center frequency of 2.8 kHz was guided by the requirement of monotonically increasing ILDs with increasing azimuth at the center frequency within the desired stimulus azimuthal range of ±70.2° to avoid ambiguous ILDs which would occur if the same cue value corresponded to multiple azimuths. Additionally, the chosen frequency range of 2-4 kHz lies in between frequency ranges that are typically either ITD- or ILD-dominant. It was therefore assumed that neither of the two binaural cues is weighted particularly strongly and each binaural cue weight has thus the potential to be increased. For this spectral range, ITD cues are conveyed mainly via the stimulus' temporal envelope, whereas ITD cues conveyed via the carrier signal (the so-called temporal fine structure) are probably only sparsely accessible through leakage into the low-frequency channels sensitive to temporal fine-structure information. Note that the particular contribution of envelope or fine-structure ITD cues was not the focus of this study. The stimulus duration was 500 ms, including 50-ms raised-cosine on/off ramps. The mean overall sound pressure level (SPL) across ears was 65 dB. To discourage listeners from using differences in the absolute level rather than ILDs for lateralization, the overall level was randomly roved from trial to trial within a ±2.5 dB range.

[1]Exact filter range: 1980-3960 Hz

Our study included stimuli with spatially consistent or inconsistent binaural cues. For consistent-cue conditions, the ITD and ILD cues corresponded to the same azimuth (squares in Fig. 1c-d), whereas for inconsistent-cue conditions, the ITD and ILD cues corresponded to disparate azimuths ('x' symbols in Fig. 1c-d). In the training sessions, 26 target azimuths (i.e., azimuths corresponding to the visually reinforced binaural cue) were distributed between -45° (left) and +45° (right) with a spacing of 3.6° (x-axis in Fig. 1d). This range corresponds to the field of view of the head-mounted display. In conditions with inconsistent ITD/ILD-combinations, the non-target azimuths (i.e., azimuths corresponding to the non-reinforced cue) were uniformly distributed ±25.2° around each target azimuth (columns in Fig. 1d), also with an azimuthal spacing of 3.6°, resulting in a total range of non-target azimuths from -70.2° to +70.2°. The disparity between ITD and ILD azimuths was intentionally limited (maximum of 25.2°) with the goal to avoid the perception of split images (e.g., Hafter and Jeffress, 1968). Based on Gaik's (1993) results, we do not expect split-image perception to occur for the frequency range and binaural cue disparities used in this study. In the pre- and posttest, in which targets were not defined because there was no visual reinforcement, all combinations of ITD and ILD azimuths used for either group during training were included (Fig. 1c). The rationale behind using so many different spatial configurations of ITD and ILD was to prevent the possibility of identifying individual stimulus azimuths and then responding strategically.

*Procedure*

The general task involved indicating the perceived azimuth of an intracranial sound source via head turn. Because stimuli were relatively narrowband and contained no HRTF filtering, they were likely not externalized by the listeners.

The experimental phases are shown in Table 1. The experiment started with a *practice session* to get participants accustomed to the task and to check their lateralization ability and sensitivity to both binaural cues. It continued with a *pretest* to measure the initial ITD and ILD weights, a seven-day *training* (completed within a two-week interval) in which the target cue was reinforced, and a *posttest* to remeasure the weights after the training. The approximate duration of the sessions was 2 hours on the first day and 1 ½ hours per day on days two to seven. The choice of the number of training sessions and their duration was guided by the training regimen used in Kumpik et al. (2010).

**Practice.** The practice session (performed at the beginning of the experiment) consisted of 130 trials that differed from training trials (see training procedure below) only in the stimuli they used. The practice stimuli had either consistent ITD/ILD-combinations (to measure basic lateralization ability) or they had either the ITD or the ILD fixed at zero while the other cue corresponded to a target azimuth (to test for sensitivity to the cues in isolation). Of the 130 practice trials, 78 (3 per target azimuth) used consistent-cue stimuli, 26 (each target azimuth once) used stimuli with ILD fixed at zero and 26 (again each target azimuth once) used stimuli with ITD fixed at zero. The trials were presented in a random order while ensuring that the first 26 trials were consistent-cue combinations.

**Testing.** The pretest and the posttest were identical for the two groups and did not contain visual reinforcement (see steps 1 and 2 of Figure 2). On each trial, participants listened to the auditory stimulus while facing the reference position and then indicated its perceived azimuth via head turn and button press. When they returned to the reference position, the session continued with the next trial. A total of 446 trials were presented, namely each ITD/ILD combination shown in Figure 1c presented once. These comprised all ITD/ILD combinations included in the training phase for both groups, as the training sets differed between groups (see below). The trials were presented in a random order and after each 150 trials, participants took a short break. This paradigm for estimating binaural cue weights is based on Stecker's (2010) "open loop" and Macpherson and Middlebrooks' (2002) methods. Namely, by asking participants to lateralize stimuli containing spatially inconsistent ITD and ILD, we can infer how much each cue contributed to the azimuthal percept by comparing the response azimuth to the azimuths of the binaural cues. This method is not subject to the bias of traditional trading ratio measurements (e.g., Deatherage & Hirsh, 1959), because no cue is actively manipulated by the participants (see Lang and Buchner, 2008).

**Training.** For training, we used a lateralization procedure based on the procedure used in Majdak, Walder and Laback (2013). We adapted it by restricting it to the horizontal (azimuthal) dimension and by using the head-pointing technique. The training procedure consisted of 5 steps (shown in Fig. 2): (1) Listening to the auditory stimulus while at the reference position (initiated by pressing the button at the reference position), (2) indicating the perceived azimuth via head turn and button press, (3) after the appearance of the visual reinforcement (a rotating cube) at the target azimuth, finding and confirming the target azimuth via head turn and button press, (4) returning to the reference position and listening to the same auditory stimulus again (initiated with a button press) while the visual reinforcement is still visible, and (5) confirming the target azimuth again via head turn and button press after which the visual reinforcement disappears. Participants were instructed to remain at the reference position for the duration of the auditory stimulus (500 ms) and reminded to do so throughout the experiment. However, participants were not physically prevented from initiating head turns during sound presentation. Nevertheless, even if they did start to turn their heads before the stimulus ended, there is no reason to expect that this would confound the experimental variables under investigation. After each 65 training trials, participants took a short break.

Auditory stimuli included both inconsistent and consistent ITD/ILD-combinations, as shown in Figure 1d. The training procedure was the same for the two groups except for which cue was the target and which the non-target. Thus, for the ITD target group, ITD azimuths did not exceed ±45° and for the ILD target group, ILD azimuths did not exceed ±45°. This ensured that the visual reinforcement was visible while at the reference position, as ±45° was the field of view of the head-mounted display. A full training session consisted of 390 trials presented in a random order. On days one and seven, participants completed only half sessions (195 trials each, created by splitting a randomized item list for a full session) to prevent fatigue because the pre- and posttest were also completed on these days. On days two to six, participants completed full training sessions.

*Analysis*

The data were analyzed using MATLAB R2018b (The MathWorks, Natick, MA). Statistical analyses of results were performed using SPSS Statistics 20 (IBM, Armonk, NY). We estimated binaural cue weights separately for each participant based on a regression analysis fitted separately for each azimuth α (between 1.8° and 45° with a 3.6° spacing between azimuths) after mirroring the data across the midline to get more reliable estimates (assuming left/right symmetry). The regression model equation is as follows:

$$R(\alpha, \Delta_{ITD}, \Delta_{ILD}) = k_{ITD}(\alpha) * \Delta_{ITD} + k_{ILD}(\alpha) * \Delta_{ILD} + Q(\alpha); \quad w_{ILD} = \frac{atan\left(\frac{k_{ILD}(\alpha)}{k_{ITD}(\alpha)}\right)}{\frac{\pi}{2}} \quad \text{(Eq. 1)}$$

REWEIGHTING OF BINAURAL CUESwhere R is the participant's response azimuth in a trial with ITD and ILD corresponding to the azimuths $\alpha + \Delta_{ITD}$ and $\alpha + \Delta_{ILD}$, respectively, $k_{ITD}$ and $k_{ILD}$ are the estimated linear regression slopes at azimuth $\alpha$ (determining the individual binaural cue weight contributions), and Q is the estimated response azimuth for consistent cues corresponding to azimuth $\alpha$. Parameter $k_{ITD}$ ($k_{ILD}$) was estimated at each azimuth by considering various azimuthal offsets (from -25.2° to +25.2°) of the cue $\Delta_{ITD}$ ($\Delta_{ILD}$) while setting the offset of the other cue, $\Delta_{ILD}$ ($\Delta_{ITD}$), to zero (as a consequence, the obtained values of $k_{ITD}$ and $k_{ILD}$ are equal to those that would be obtained by performing independent univariate fits separately for $k_{ITD}$ and $k_{ILD}$). Parameter Q, on the other hand, is estimated for each azimuth $\alpha$ by all the data for which either the ITD or ILD azimuth corresponded to $\alpha$. Thus, referring to Fig. 1c, the model was fitted for each azimuth $\alpha$, indicated by a square, by considering only items of the row and column that include that square (an example set of data used for the estimation of parameters at $\alpha$ of 9° is indicated by the black frame in Fig. 1c). These estimates of $k_{ITD}$ and $k_{ILD}$, representing orthogonal vectors, were then combined to derive the ILD weight $w_{ILD}$ (note that $w_{ITD}$ = $1 - w_{ILD}$).

**Results**

Figure 3 provides an overview of the overall results pattern by plotting the mean response azimuth across participants as a function of either target azimuth (panels a-d, left-hand side) or non-target azimuth (panels e-h, right-hand side). The data is parameterized by the offset of the cue not shown on the x-axis from the cue shown on the x-axis, pooling across three offset ranges: central offsets (-3.6°, 0°, and 3.6°, green circles), leftward offsets (≤14.4°, blue downward-pointing triangles) or rightward offsets (≥14.4°, red upward-pointing triangles). The response curves are fairly linear and parallel to the diagonal, showing that participants were able to extract the binaural cues and responded accurately using the employed setup. The separation of the three lines in each panel shows that both cues contributed to the perceived azimuth (e.g., response azimuths are further right for rightward offsets compared to central offsets, independent of which cue is parameterized as shown by the red lines falling above the green lines). In the posttest, response slopes were generally flatter compared to the pretest, suggesting an overall compression of response azimuths which is further explored below. The distance between the three curves is indicative of binaural cue weighting. It is larger in conditions where the ITD offset is shown as the parameter, suggesting a larger ITD weight in both subject groups, which is particularly prominent in the pretest data (panel c versus a and panel e versus g). For panels showing the non-target offset as the parameter (a-d), the distance between curves is expected to be smaller in the post- compared to the pretest, assuming a reduction of the non-target cue weight and thus an increase of the target cue weight. This indeed seems to be the case for both groups. Note, however, that compression (i.e., overall flatter slopes in the posttest) also contributes to a reduced distance between curves. In contrast, for panels showing the target offset as the parameter (e-h), binaural cue reweighting would result in increasing distance between curves from pre- to posttest. In this case, however, compression counteracts cue reweighting. Consistent with these assumptions, the distance between curves appears to be more similar in pre- and posttest.

Figure 4 shows the ILD weights ($w_{ILD}$) as determined by the regression analysis at each azimuth separately for the two groups. As hypothesized, posttest ILD weights (red squares) *decreased* for the ITD target group and *increased* for the ILD target group compared to pretest ILD weights (blue circles). A 2 x 13 x 2 mixed-design ANOVA with the within-participants factors *time* (pre- vs. posttest) and *azimuth* and the between-participants factor *group* (ITD vs. ILD target group) yielded no significant main effects, but a significant *time* x *group* interaction ($F(1,18) = 20.44$, $p < .001$, $\eta_p^2 = .532$). All other interactions were non-significant. The lack of a significant main effect of *time* was expected, assuming opposing effects of time on the ILD weights for the two groups. Therefore, and given the significant *time* x *group* interaction, follow-up partial ANOVAs were performed separately for the two groups. For the ITD target group, there was a significant main effect of *time* ($F(1,9) = 7.47$, $p = .023$, $\eta_p^2 = .454$) with larger ILD weights in the pretest ($M = 0.40$, $SD = 0.13$) compared to the posttest ($M = 0.31$, $SD = 0.09$), while neither the main effect of *azimuth* ($F(12,108) = 1.65$, $p = .088$, $\eta_p^2 = .155$) nor the *time* x *azimuth* interaction was significant ($F(12,108) = 0.95$, $p = .500$, $\eta_p^2 = .096$). Similarly, for the ILD target group, the ANOVA yielded a significant main effect of *time* ($F(1,9) = 13.39$, $p = .005$, $\eta_p^2 = .598$) with smaller ILD weights in the pre- ($M = 0.31$, $SD = 0.09$) compared to the posttest ($M = 0.43$, $SD = 0.10$), while neither the main effect of *azimuth* ($F(12,108) = 1.17$, $p = .314$, $\eta_p^2 = .115$) nor the *time* x *azimuth* interaction ($F(12,108) = 1.34$, $p = .205$, $\eta_p^2 = .130$) was significant. To check whether pretest differences between the groups could have driven the effect, we additionally ran two separate 2 *(group)* x 13 *(azimuth)* mixed-design ANOVAs, one for the pre- and one for the posttest data. While the main effect of *group* was not significant in the pretest ($F(1,18) = 3.17$, $p = .092$, $\eta_p^2 = .150$), there was a significant group difference in the posttest ($F(1,18) = 8.04$, $p = .011$, $\eta_p^2 = .309$). Note that the estimates of $w_{ILD}$ are independent of response compression under the assumption that compression affects the target and non-target cue equally, so that its effect on the slopes $k_{ITD}$ and $k_{ILD}$ cancels out in the final weight estimation (Eq. 1). This assumption was tested using a modeling approach, which yielded a better account of the data for a model version assuming reweighting combined with cue-independent compression than a model version assuming cue-specific compression of the non-target cue (see appendix for details). Taken together, these results suggest that the training induced an azimuth-independent increase in target-cue weights for both groups.

Next, we addressed the potential implication of apparently asymmetric binaural cue weights. Since azimuth had no significant effect on the ILD weights, the weights were averaged across azimuths. The mean ILD weights were significantly smaller than 0.5 (a weight of 0.5 means equal weighting of the two binaural cues) in both the pretest ($M = 0.35$, $SD = 0.12$, $T(19) = -5.59$, $p < .001$, $d_z = 1.25$) and the posttest ($M = 0.37$, $SD = 0.11$, $T(19) = -5.37$, $p < .001$, $d_z = 1.20$), showing an overall dominance of ITD cues. Because a higher baseline (pretest) weight for the target cue potentially limits the room for training-induced reweighting towards the target cue, we first checked if the amount of reweighting, quantified as the post- versus pretest difference in target cue weights, differed between groups. The amount of reweighting (ITD target group: 0.09; ILD target group: 0.12) did in fact not differ significantly between groups ($T(18) = -0.66$, $p = .516$, $d = -0.31$), providing no evidence that such ceiling effects might have affected the ITD target group. As the distribution of pre- and posttest weights across listeners might provide further hints, Figure 5 plots the posttest target cue weight as a function of the pretest target cue weight. For the ITD target group (left panel), the data points accumulate more towards the upper right part of the plot and the pattern appears to be flatter. We therefore cannot completely role out that listeners with a high pretest ITD weight were somewhat affected by ceiling effects. For the ILD target group (right panel) the data accumulate more towards the lower left part of the plot and the pattern is more parallel to the diagonal. Thus, there is no hint for ceiling effects in the ILD target group, which is expected given the low pretest ILD weight of all listeners.

We further sought to investigate the time course of cue reweighting across training sessions. However, our regression analysis required a balanced distribution of ±25.2° of ITD azimuths around ILD azimuths and vice versa, which was available only for azimuths up to 19.8°, as target azimuths were limited to ±45° during training. Note that during the pre- and posttest a balanced distribution was provided up to ±45° for both target and non-target cues. We therefore gradually limited the range of $Δ_{ITD}$ and $Δ_{ILD}$ with increasing azimuth for azimuths more lateral than 19.8°, always ensuring a balanced, albeit reduced distribution (for example, at 41.4°, only the two neighboring azimuths as well as the consistent-cue condition were considered, see Fig. 1d). To compare the training sessions to the pre-and posttest, this modified analysis was also performed for the pre-/posttest data. Responses were averaged across azimuths before running the regression analysis, since the model could not be fitted for all azimuths and time points separately, as in training sessions one and seven, only half of the items were presented. Figure 6 shows the training progress for this data subset for the two subject groups. For comparison, the pre- and posttest weights calculated using all data, replicated from the means across azimuth in Figure 4, are added as separate filled symbols at the far left and far right of the figure. For both groups, the training effect seems to have been induced within the first training session and there appears to be no further change over the course of training. Accordingly, there was no significant effect of *time* in either repeated-measures ANOVA (separate for the groups), using the seven training sessions as the within-subjects factor. Since there was no significant difference between training sessions, we averaged across training sessions to further explore the relationship between the pre-/posttest and the training. We ran two repeated-measures ANOVAs (one for each group) with the within-subjects factor *time* (pretest – averaged training sessions – posttest) including post-hoc pairwise comparisons. For the ITD target group, the main effect of *time* failed to reach significance after correcting for a sphericity violation ($F(1.27,11.44) = 3.72$, $p = .072$, $\eta_p^2 = .292$, Greenhouse-Geisser-corrected). The post-hoc pairwise comparisons also did not yield any significant differences (all $p > .191$, Bonferroni-corrected). For the ILD target group, the repeated-measures ANOVA yielded a significant main effect of *time* ($F(1,18) = 19.55$, $p < .001$, $\eta_p^2 = .685$). The post-hoc pairwise comparisons showed significant differences between the pretest and the training ($p = .001$, Bonferroni-corrected) with smaller ILD weights in the pretest ($M = .315$, $SD = .079$) compared to the training ($M = .594$, $SD = .089$), the training and the posttest ($p = .022$, Bonferroni-corrected) with larger ILD weights in the training compared to the posttest ($M = .405$, $SD = 092$) as well as the pre- and the posttest ($p = .048$, Bonferroni-corrected) with smaller ILD weights in the pretest compared to the posttest. Hence, the ITD target group showed a modest, albeit non-significant decrease in the ILD weight from the pretest to the first training session and no change afterwards. In contrast, the ILD target group showed a larger, significant increase in the ILD weight during the first training session and a significant reduction of the ILD weight from the last training session to the posttest. Note that these training results should be interpreted with caution, given that only a subset of data could be analyzed, given the requirement to estimate binaural cue weights independent of response bias.

Next, we examined the response compression from pre- to posttest observed for both groups (see description of Fig. 3). Figure 7 shows the estimated response azimuths for consistent cues based on the regression analysis (i.e., the estimated values of Q from Eq. 1), pooled across groups. Lateralization in the pretest was fairly accurate, since the estimated response azimuths are similar to the cue azimuths (i.e., close to the diagonal; mean pretest slopes of estimated response vs. cue azimuths for the ITD and ILD target group: 1.06 and 1.02, respectively). But there is a flattening of the slope from pre- to posttest suggesting a systematic and apparently linear compression of responses (mean posttest slope for both the ITD and ILD target groups was 0.87). Subjecting the ratio of estimated responses/cue azimuths to a 2 *(time)* x 13 *(azimuth)* x 2 *(group)* mixed-design ANOVA yielded a significant main effect of *time* ($F(1,18) = 18.19$, $p < .001$, $\eta_p^2 = .503$), but neither the main effects (*azimuth* and *group*) nor the interactions were significant, suggesting that this compression of responses from pre- to posttest is indeed linear and similar in both groups.

To evaluate the overall lateralization performance in our experiment, we calculated a basic, widely used measure, the root-mean-square error (RMSE) between response and stimulus azimuth for the consistent-cue items. The RMSE contains errors due to both systematic bias and response variability. In the pretest, the mean across all participants was 13.04° (*SD* = 4.94). Using a head pointing method in a virtual environment including individual HRTF-filtering, Middlebrooks (1999) reported a mean lateral RMSE of 14.5° (*SD* = 2.2) and Majdak, Goupell and Laback (2010) a mean lateral RMSE of 14.4° for untrained listeners. Our lateralization method using binaural stimuli without HRTF filtering therefore shows comparable accuracy to virtual acoustics studies using a similar paradigm.

Finally, we attempted to quantify the training-induced change in overall lateralization precision (i.e., the consistency in lateralization responses; see Heffner & Heffner, 2005) and confirm that the listeners did not perceive split images (Hafter & Jeffress, 1968) when stimuli with inconsistent cues were presented. To that end, response variability was calculated by computing the residuals (where, for each response, the residual is defined as the deviation of the actual response azimuth from the response azimuth predicted by the regression analysis from Eq. 1) and then computing the standard deviation of these residuals. Figure 8 shows the results as a function of cue disparity, averaged across groups. The response variability was systematically lower in the posttest than in the pretest, while there seems to be no systematic effect of cue disparity. The mean variability across groups and all binaural cue disparities decreased from 10.62 (*SD* = 3.90) in the pretest to 6.54 (*SD* = 2.17) in the posttest. A 2 (*time*) x 8 (*cue disparity*) x 2 (*group*) mixed-design ANOVA showed significant main effects of the factors *time* ($F(1,18) = 27.24$, $p < .001$, $\eta_p^2 = .602$) and *cue disparity* ($F(4.18, 75.26) = 2.51$, $p = .046$, $\eta_p^2 = 123$, Greenhouse-Geisser-corrected) but no significant effect of the factor *group* and no significant interactions. Post-hoc pairwise comparisons revealed that the main effect of *cue disparity* was driven by a significant difference between cue disparities of 14.4° and 25.2° ($p = .008$, Bonferroni-corrected). However, variability did not systematically increase with increasing cue disparity nor adopted an inversed u-shape. Either of these two patterns might be expected if larger cue disparities had evoked the perception of split images. As a further check for the possibility of split image perception, we inspected the response distributions as a function of binaural cue disparity for each participant and found no systematic indications for distributions being bimodal or centered close to one cue azimuth only, as may be expected in case of split image perception. Thus, these results provide no indication for split image perception and instead suggest that participants perceived a single compact auditory image for all cue disparities included in this study.

**Discussion**

This study is, to our knowledge, the first one to provide evidence that it is possible to increase the relative weighting of both the ITD and the ILD cues. This binaural cue reweighting was induced through a lateralization training in a virtual audio-visual environment, employing visual reinforcement of the target cue as well as symmetric azimuthal variation of non-target cues around target cues.

While several studies addressed the plasticity of the spatial auditory system to spatial cue modifications (e.g., Kumpik et al., 2010; Shinn-Cunningham et al., 1998), the only two published studies we are aware of explicitly addressing binaural cue reweighting (Jeffress & McFadden, 1971; Kumpik et al., 2019) did not produce conclusive results. Jeffress and McFadden observed no reweighting effect and while Kumpik et al. report an increase in ILD weighting after ITDs were randomized during a visual oddball task, they observed an even stronger increase in ILD weighting when spatially consistent ITDs and ILDs were presented. Furthermore, they did not observe an increase in ITD weighting when ILDs were randomized, making it difficult to draw strong conclusions. These different results can likely be attributed to differences in the methodology. Jeffress and McFadden used a discrimination task in which participants indicated whether they perceived the auditory stimulus left or right from the midline. This task provides relatively little information compared to the lateralization task used here, in which the perceived azimuth was used as a measure. Additionally, their auditory stimuli were in a frequency range where only ITDs but not ILDs arise naturally, their auditory and visual stimuli were not presented simultaneously, and their paradigm provided no proprioceptive feedback. Therefore, their null result is to be expected, for example, if bottom-up multisensory integration is required to induce reweighting. Finally, due to the small sample size of only four participants, their study might not have had enough power to detect an effect. In Kumpik et al.'s study, the auditory stimuli were not needed to perform the training task and thus might not have received enough attention. Furthermore, they applied reverberation, presumably making the ITDs less reliable, which is in line with Rakerd and Hartmann's (2010) observation that responses follow ILDs in reverberant environments. This may have restricted the potential to increase the ITD weighting.

    Our stimuli were presented via headphones without being HRTF-filtered, and, thus, they did not convey monaural spectral localization cues that are potentially informative about the azimuth of the stimulus. The pre-/posttest differences observed in the present study can therefore not be attributed to increased weighting of monaural cues, as in Kumpik et al. (2010), in which the authors did not observe any adaptation to the changed binaural cues.

    Although our data is not well suited for studying the time course of reweighting (e.g., not all ITD/ILD combinations included in the pre- and posttest were included in the training), it seems reasonable to conclude from the current analysis of the training data that reweighting occurred

predominantly within the first training session. Interestingly, our results differ in the time course from those of Kumpik et al. (2010), who did not observe adaptation when all training trials were performed in one day, but rather found continuous improvement across all 7 or 8 training days. A plausible explanation could be that the listeners in their study needed to learn to exploit monaural spectral cues for azimuthal localization when the binaural cues were disrupted, while such learning was not required in our study given that both binaural cues are used by default. Accordingly, our observation that adaptation occurred within the first training session is similar to Kumpik et al. (2019), who reported binaural cue reweighting to occur within less than 1 hour of training. It should also be noted that the time course seems to differ between the two groups in the current study. While the ITD target group showed less reweighting from the pretest to the training which then remained stable through the posttest, the ILD target group showed strong reweighting from the pretest to the training, part of which then got lost from the last training session to the posttest. Wright and Fitzgerald (2001) for example also report different time scales for ITD vs. ILD discrimination learning. However, these time scales differ from the current observations. Namely, they report an initial rapid improvement for both cues that generalizes across conditions followed by a slower improvement for ILD discrimination only. Another possibility to consider is that the adaptation occurs on multiple time scales, as, for example, observed in the ventriloquism aftereffect (Bosen, Fleming, Allen, O'Neill, & Paige, 2018). Specifically, the reweighting of the ILD target group might consist of a quick strong component that is spontaneously reversed when the visual feedback stops, combined with a weaker, but more sustained reweighting. In comparison, the ITD target group might only show the more sustained component. Taken together, there is not enough evidence to draw strong conclusions about the time course of binaural-cue reweighting, making this an important topic for future studies.

    The ILD weights in our study were found to be significantly smaller than 0.5 in both pre- and posttest. This is consistent with Macpherson and Middlebrooks' (2002) results. Our stimuli are best comparable to their wideband stimuli, since our passband lies in the range between their low-pass

and high-pass filter conditions and some low- and high-frequency energy was present due to the roll-off of our band-pass filter. For Macpherson and Middlebrooks' wideband stimuli, the listeners either weighted ITDs and ILDs equally or ITDs were weighted more, which compares well to our present results. The observed differences in the pretest weights of ITD and ILD could have resulted in a different task difficulty between the two groups. Since ITDs were weighted more strongly already in the pretest, ceiling effects might have limited the potential for reweighting for the ITD target group. Although the group means did not differ in the amount of reweighting, members of the ITD target group tended to show decreasing reweighting with increasing baseline (pretest) weight, potentially indicating ceiling effects for some listeners in this group.

We found a systematic compression of the response azimuth range from pre- to posttest. This can at least partly be attributed to limiting the target azimuth range to ±45° during training. While the non-target cues ranged up to ±70.2° and auditory stimuli were certainly perceived beyond ±45°, target azimuths and therefore also the visual reinforcement were restricted to ±45°, likely triggering a mapping to this azimuth range. Interestingly, the resulting response compression occurred not only at the edges, but across the entire azimuth range, and followed a linear function. This result is consistent with an earlier study showing that various nonlinear mapping functions between auditory and visual azimuth space resulted in response azimuths following a linear approximation of these functions (Shinn-Cunningham et al., 1998). Furthermore, fatigue or decreased willingness by the participants to exploit the whole azimuthal response range in the posttest might have contributed to response compression. We considered the theoretical possibility that response compression was specific to the non-target cue, given that the non-target cue was varied over a larger range than the target cue during training. This would undermine the assumption of our final weight estimation based on the regression analysis that compression equally affected the ITD and ILD cues. We tested this possibility by predicting the posttest data with two versions of a model, assuming either cue-specific compression of the non-target cue only or cue reweighting combined with cue-independent compression(see Appendix). The first model version provided a REWEIGHTING OF BINAURAL CUESsignificantly better account of the data, supporting our hypothesis that the training resulted in increased target-cue weight, in addition to overall (cue-independent) compression.

As a result of the extensive training, participants showed substantial improvement in lateralization precision, as shown by the significantly reduced response variability in the posttest. Such improvement can be due to procedural training (i.e., to more focused attention or more precise use of the equipment) and to overall perceptual training (Hawkey, Amitay, & Moore, 2004). Importantly, we argue that these effects are independent of cue reweighting, since we used the standard deviation of the residuals of the regression analysis as a measure of response variability, which thus accounts for the effects of reweighting and response compression.

An important question is whether the current results could have been mediated by conscious or strategic listening. We consider strategic reweighting very unlikely given the myriad of target and non-target azimuths spread across a wide range. Moreover, the disparity between ITD and ILD azimuths was intentionally limited (maximum of 25.2°) with the goal to avoid the perception of split images (e.g., Hafter and Jeffress, 1968) which could theoretically allow for strategic listening, if the two binaural cues could be distinguished. In fact, all listeners (as well as the authors during informal piloting) reported to perceive compact auditory images, with no indication of split images. Consistent with these subjective reports, we observed similar response variability across cue disparities as well as no indications for response distributions being bimodal or centered close to the azimuth of one binaural cue only.

We aimed at maximizing chances for binaural cue reweighting to demonstrate the general feasibility of inducing such an effect. To that end, we manipulated the stability of the cues by varying the non-target cue over a larger range than the target cue and presented visual reinforcement at the target azimuth after the response as top-down feedback as well as simultaneously with the second sound presentation to tap into multisensory bottom-up processes. Therefore, we cannot disentangle the individual contributions of each of these factors to binaural cue reweighting, a question that could be addressed in follow-up studies. Furthermore, we intentionally chose a frequency range that lies in between typically ITD- or ILD-dominant regions so that both ITD and ILD weighting had the potential to be increased. With these stimuli, however, it is unclear if or to what extent fine-structure ITD cues may have perceptually contributed in addition to envelope ITDs. Future studies are needed to separate possible contributions of fine-structure versus envelope ITD cues to binaural cue reweighting. Additionally, since we trained and tested with 2-4 kHz noise bursts and used a lateralization task, it is unclear whether the training-induced reweighting generalizes to non-trained frequency regions or to other tasks such as trading ratio measurements (e.g., Deatherage & Hirsh, 1959) or discrimination tasks (Jeffress & McFadden, 1971), which would be important for judging the ecological relevance of the phenomenon. Wright and Fitzgerald (2001), for example, observed different generalization patterns of ITD and ILD sensitivity training. Additionally, it is of interest whether the changed weights can affect speech understanding in multi-talker environments. For example, since listeners benefit from the presence of ITD information in spatial release from speech masking (e.g., Ellinger, Jakien, & Gallun, 2017; Kidd, Mason, Best, & Marrone, 2010), increased ITD weights could be associated with better speech understanding in multi-talker environments where salient ITD cues are available.

Finally, our results might have interesting implications for listeners with cochlear implants (CIs). It has been shown that localization in the horizontal plane with current CI systems is almost entirely based on ILDs, while ITDs contribute only very little or not at all (Grantham, Ashmead, Ricketts, Haynes, & Labadie, 2008; Seeber & Fastl, 2008). On one hand, this is due to the properties of current envelope-based CI systems, which encode no useful ITDs in the pulse carriers and whose envelope ITD cues for real-life stimuli are not very salient (Grantham et al., 2008; Laback, Pok, Baumgartner, Deutsch, & Schmid, 2004; Laback, Zimmermann, Majdak, Baumgartner, & Pok, 2011). On the other hand, even when presenting pulse carrier ITD highly controlled via a research system at a single interaural electrode pair, CI listeners' sensitivity is greatly reduced and much more variable across listeners compared to normal-hearing listeners' carrier ITD sensitivity (Laback, Majdak, & Baumgartner, 2007; Majdak, Laback, & Baumgartner 2006; van Hoesel, 2007). Several explanations have been proposed for this perceptual deficit in electric hearing (see, e.g., Laback, Egger, and Majdak, 2015). Considering our current results, it might partly be a result of reweighting of the binaural cues over time. Specifically, it is possible that binaural cue reweighting takes place after CI implantation, resulting in a stronger weighting of the ILDs which consistently indicate sound source locations, and a decreased weighting of the ITDs which are not reliably and saliently provided by the CI listeners' clinical devices.

In conclusion, our results suggest that reweighting of auditory localization cues is not limited to monaural vs. binaural cues, which has been shown in previous studies, but that reweighting of the two binaural cues ITD and ILD is also possible. Specifically, we show that both the ITD and ILD weighting can be increased by reinforcing the respective cues through lateralization training. This could play a role in adapting to variable acoustic environments, be a factor contributing to the low contribution of ITDs to sound localization of CI listeners, and have potential applications, for example, in training for unfamiliar audio-visual environments or hearing devices that impede one of the two binaural cues.

**Acknowledgments**

This research was supported by the OeAD, project SpaCI (#MULT_DR 11/2017), the uni:docs Fellowship Program for Doctoral Candidates of the University of Vienna, and the SRDA, project DS-FR-19-0025.

**Appendix**

The estimate of $w_{ILD}$ is independent of response compression under the assumption that compression affects the target and non-target cue equally, so that its effect on the slopes $k_{ITD}$ and $k_{ILD}$ cancels out in the final weight estimation (Eq. 1). However, as non-target azimuths were varied over a wider range (±70.2°) than target azimuths (±45°) during training, it is conceivable that the training led to a specific compression of the non-target cue while the target cue was not affected by compression. Such a scenario could be an alternative interpretation of the results because it would reduce the relative contribution of the non-target cue, similar to the effect of cue reweighting. In this Appendix we describe the results of a simple modeling approach to determine which of these two scenarios quantitatively better characterizes the current data. To that end, we predicted lateralization responses based on weighted averages of the cue azimuths under different compression assumptions and compared those predictions to the actual posttest data. The mean posttest data of each group from Fig. 3 were used, for which either the non-target cue was varied around the target cue (panels b and d) or the target cue was varied around the non-target cue (panels f and h), categorized by cue offsets "center", "left" or "right" (see caption of Fig. 3). Predictions were made separately for each azimuth.

For a given condition with a certain combination of binaural cues values, corresponding cue azimuths were first multiplied by a compression factor, C (C=1: no compression: C<1: compression), which was either the same (cue-independent) or different (cue-specific) for the two cues (see model versions below). The resulting values where then subjected to weighted averages as determined by the parameter $w_{ILD}$ (see description of regression analysis in the methods section) to form the predicted response azimuth. Depending on the model version, either or both parameters C and $w_{ILD}$ were freely varied (in steps of 0.01) and the parameter(s) resulting in the highest prediction accuracy were determined. Prediction accuracy was determined by means of the RMS deviation between observed and predicted response azimuths (referred to as RMSE). All conditions fulfilled the requirements for convex optimization.

We compared the performance of two model versions. In the *cue-specific-compression* (*CSC*) model, C was freely varied for the non-target cue while it was fixed for the target cue. Specifically, for the target cue, C was based on the consistent-cue pretest data (1.070 and 1.024 for the ITD and ILD target groups, respectively). Parameter $w_{ILD}$ was fixed and taken from the regression analysis of the pretest data, where C is very close to 1 and, thus, weight estimation is very unlikely to be confounded by cue-specific compression. In the *reweighting-and-cue-independent-compression* (*RCIC*) model, both C and $w_{ILD}$ were freely varied, i.e., all combinations of the two parameters were evaluated.

The results are summarized in Table A1. The RCIC model resulted in a lower RMSE (1.09°) than the CSC model (1.54°). This difference was significant for both the ITD target group ($T(12)$ = 3.57, $p$ = .004, $d_z$ = 0.99) and the ILD target group ($T(12)$ = 3.09, $p$ = .006, $d_z$ = 0.86) when comparing the means across azimuths. Moreover, the estimates of $w_{ILD}$ across azimuths by the RCIC model correspond very well to the respective posttest estimates by the regression analysis ($r^2$ = 0.9 and 0.94 for the ITD and ILD target groups). Also, mean C estimates by the RCIC model (0.89 and 0.87 for ITD and ILD target groups, respectively) correspond very well to the bias estimates of the regression model (0.87 and 0.87 for ITD and ILD target groups, respectively). In summary, the model results suggest that the training effect observed in the posttest was more likely induced by cue reweighting combined with cue-independent compression than by cue-specific compression of the non-target cue.

| Model version | Compression | Cue weight | Mean RMSE across azimuths | SD | C | SD | $w_{ILD}$ | SD |
|---|---|---|---|---|---|---|---|---|
| Cue-specific compression (CSC) | varied | pretest | 1.54 | 0.55 | 0.70 | 0.10 | - | - |
| Reweighting and cue-independent compression (RCIC) | varied | varied | 1.09 | 0.36 | 0.89 | 0.03 | 0.35 | 0.03 |
| Cue-specific compression (CSC) | varied | fixed (pretest) | 2.37 | 0.85 | 0.79 | 0.03 | - | - |
| Reweighting and cue-independent compression (RCIC) | varied | varied | 1.96 | 0.60 | 0.87 | 0.03 | 0.45 | 0.06 |

Table A1. *Results of modeling the posttest data for the ITD target group (upper two rows) and ILD target group (lower two rows) using two model versions, the Cue-Specific Compression (CSC) model and the Reweighting and Cue-Independent Compression (RCIC) model. RMSE refers to the* RMS deviation between observed and predicted response azimuths, *C to the compression factor, and* $w_{ILD}$ *to the ILD cue weight.*

| Day 1 | Day 2 | Day 3 | Day 4 | Day 5 | Day 6 | Day 7 |
|---|---|---|---|---|---|---|
| Practice (130 trials) | | | | | | |
| *Pretest* (446 trials) | | | | | | |
| Training (195 trials) | Training (390 trials) | Training (390 trials) | Training (390 trials) | Training (390 trials) | Training (390 trials) | Training (195 trials) |
| | | | | | | *Posttest* (446 trials) |

Table 1. *Time course (top to bottom) of experimental phases. On day 1, participants completed the practice session, followed by the pretest, followed by half a training session. On days 2-6, participants completed one training session each. On day 7, participants completed half a training session followed by the posttest.*

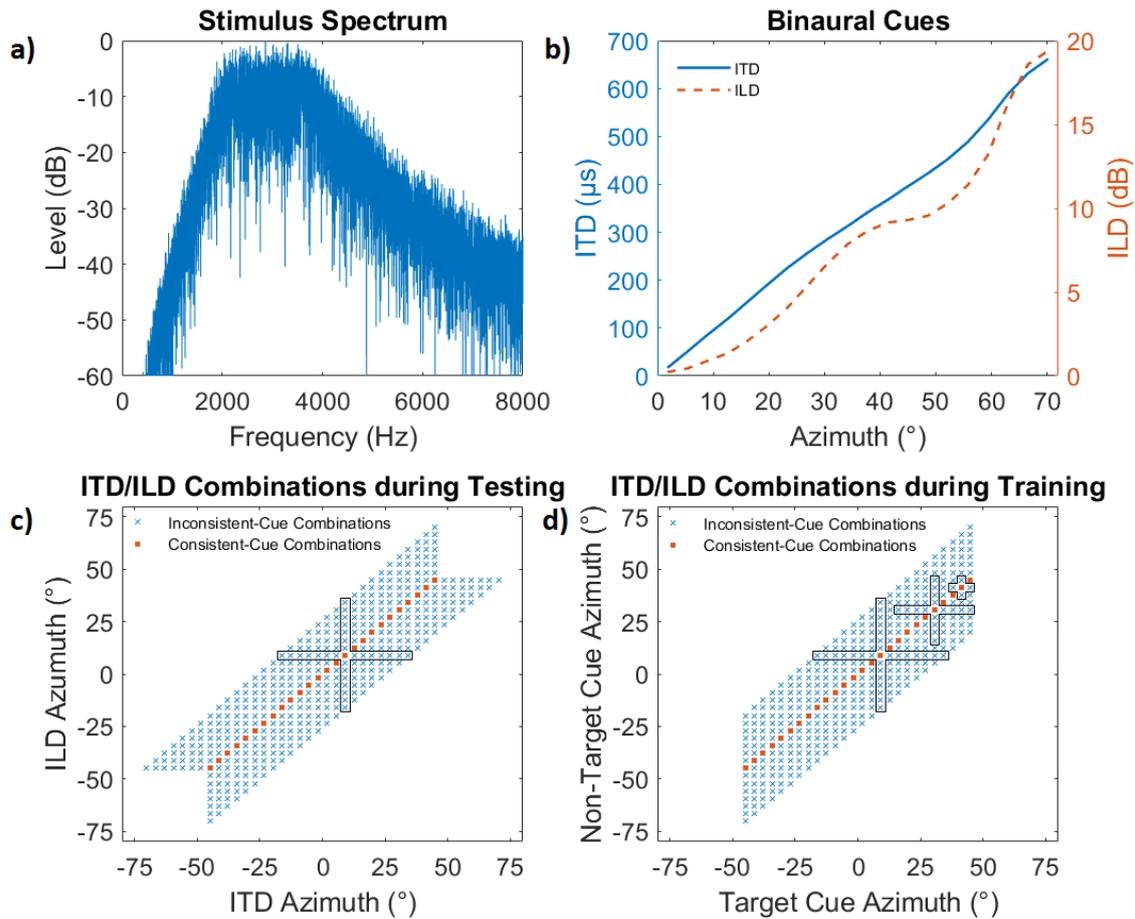

Figure 1. *Experimental setup and stimuli. Panel a) shows the spectrum of the auditory stimuli used. Panel b) shows the functional relation between the azimuth and the binaural cues as derived by Xie (2013) from the KEMAR head-related impulse responses (HRIRs), with ITDs (solid line) referring to the left ordinate and ILDs (dashed line) referring to the right ordinate. ITDs are based on broadband cross-correlation of the left and right ear HRIRs. ILDs are based on HRTF magnitudes at 2.8 kHz. Panel c) shows all the ITD/ILD-azimuth combinations used in the pre- and posttest. The frame indicates the data that were used to estimate the model parameters for the pre-/posttest data at one example azimuth (9°). Panel d) shows all target and non-target cue combinations used in the training. For the ITD target group, target and non-target cues were ITD and ILD, respectively, and for the ILD target group, target and non-target cues were ILD and ITD, respectively. The frames indicate the data that were used to estimate the model parameters for the training data at example azimuths of 9° (large frame), 30.6° (medium frame), and 41.4° (small frame). The reduction of frame size towards the edge was required to ensure symmetric distributions (see text).*

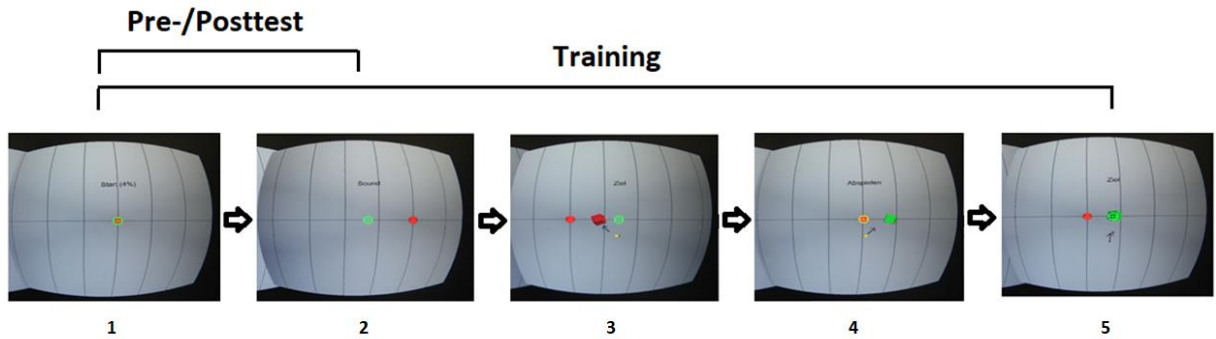

Figure 2. *Time course of a trial during testing (panels 1-2) and (pre-)training (panels 1-5). 1) Participants oriented towards the reference position (indicated by a red sphere) by turning their head (guiding a green crosshair) towards it and pressed a button to elicit the sound presentation. 2) Participants turned their head to the perceived azimuth and pressed a button (in this example, they performed a head-turn to the left). 3) Visual reinforcement (a rotating red cube) appeared at the target azimuth. The participants confirmed the target azimuth via a head turn to the target azimuth and a button press. 4) The visual reinforcement turned green, and participants returned to the reference position. After another button press, the auditory stimulus was presented again while the visual reinforcement was still visible. 5) Participants performed another head turn to the target azimuth and pressed the button. In step 3 and 5, the button-press was accepted only if the head-orientation (green crosshair) was within ±5° of the target.*

REWEIGHTING OF BINAURAL CUES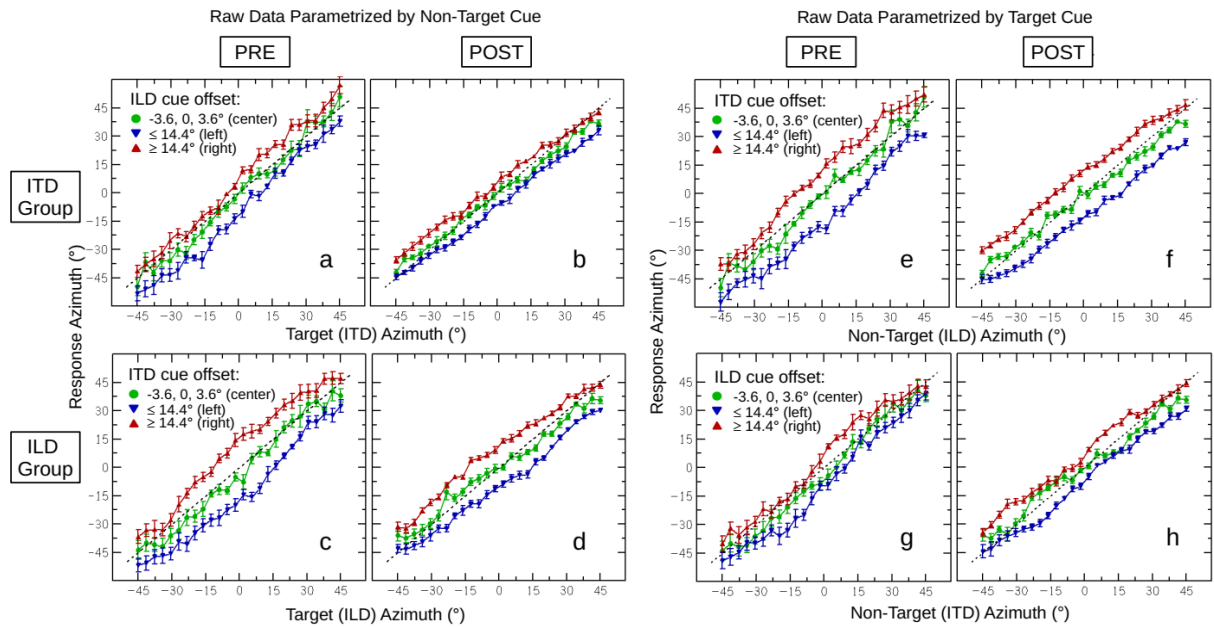

Figure 3. *An overview of mean response azimuths across listeners as a function of either target azimuth (ITD azimuth for the ITD target group, ILD azimuth for the ILD target group; panels a-d, left-hand side) or non-target cue azimuth (ILD azimuth for ITD target group, ITD azimuth for ILD target group; panels e-h, right-hand side). The data is parameterized by the offset of the cue not shown on the x-axis from the cue shown on the x-axis, pooling across three offset ranges: central offsets (-3.6°, 0°, and 3.6°, green circles), leftward offsets (≤14.4°, blue downward-pointing triangles) or rightward offsets (≥14.4°, red upward-pointing triangles). In each panel pair, the panel on the left shows the pretest and the panel on the right shows the posttest. Error bars indicate standard errors of the mean.*

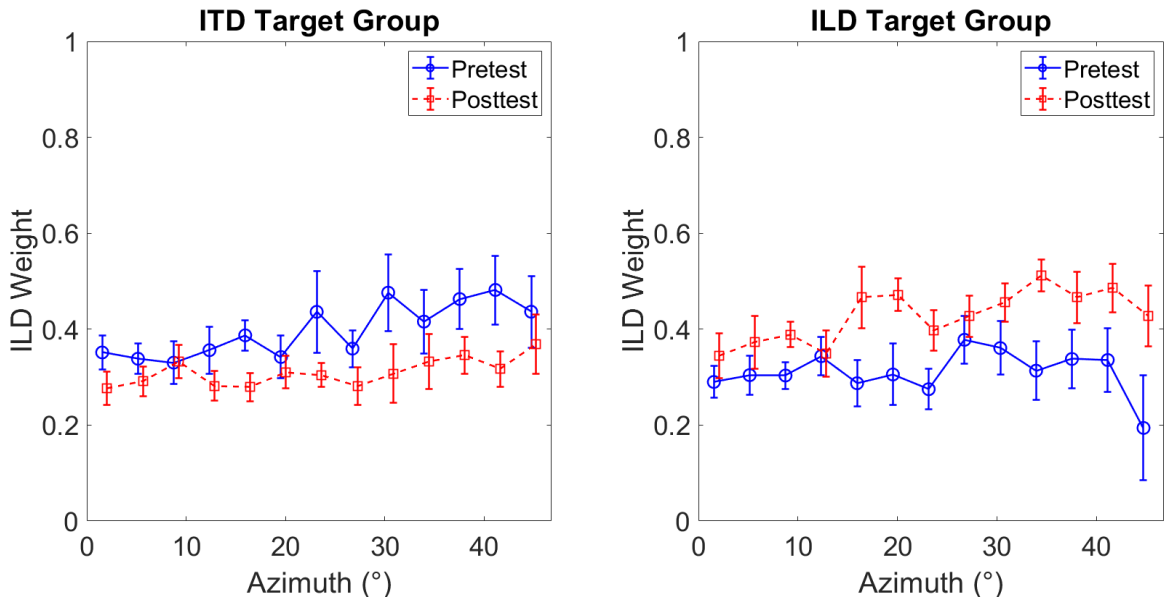

Figure 4. *ILD weights as a function of azimuth (after collapsing data across left and right azimuths) derived using a regression analysis. Blue circles show the pretest results and red squares the posttest results, separately for the ITD and ILD target groups (individual panels). Note that, by definition, ITD weight = 1 – ILD weight. Error bars show standard errors of the mean.*

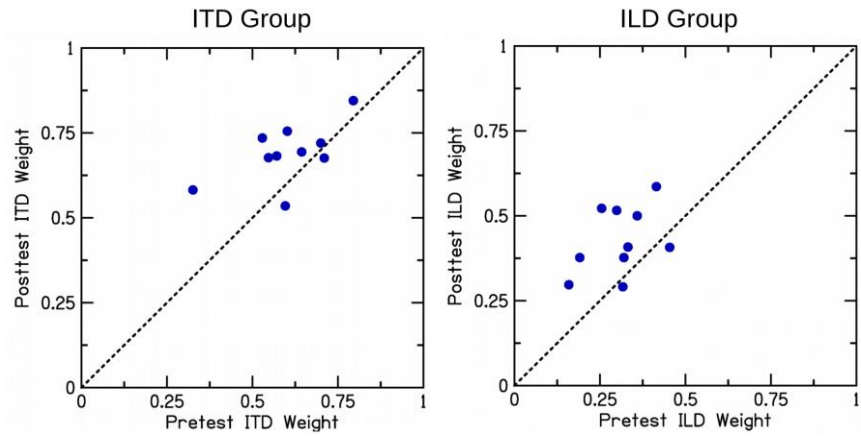

Figure 5. *Posttest target cue weight as a function of pretest target cue weight. The symbols show individual participants. For the ITD target group (left panel), the data points accumulate more towards the upper right part of the plot and the pattern appears to be flatter compared to the pattern of the ILD target group (right panel), whose data accumulates more towards the lower left part of the plot.*

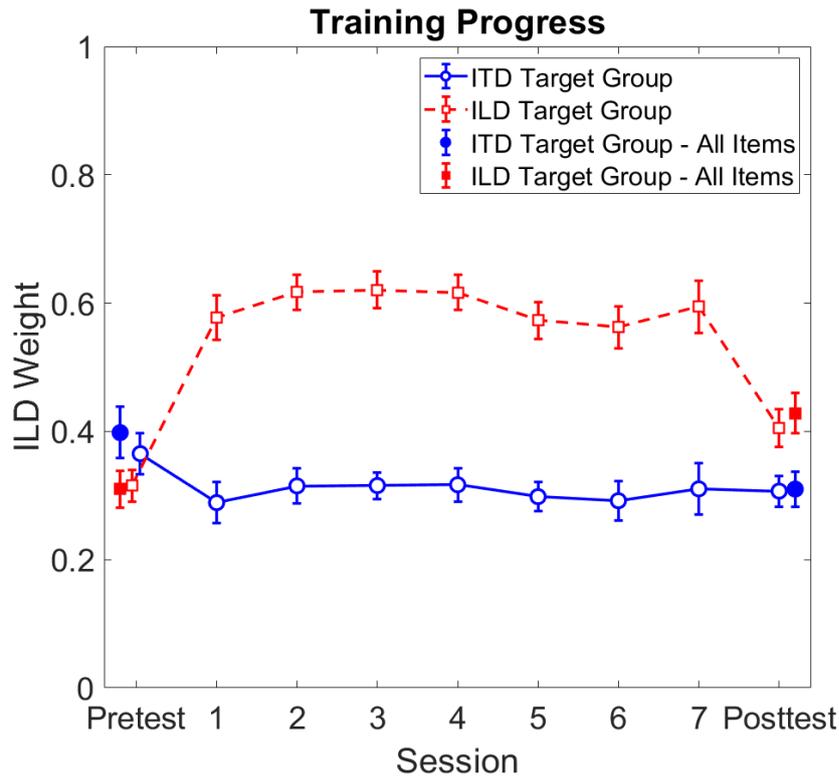

Figure 6. *Training progress. ILD weights calculated by limiting the combinations considered for azimuths more lateral than 19.8° to ensure balanced distributions are shown for the pretest, the seven training sessions and the posttest. Blue circles show the results for the ITD target group, red squares for the ILD target group. For comparison, the pre- and posttest weights calculated using all data are shown as separate symbols at the far left and far right of the figure. Error bars show standard errors of the mean.*

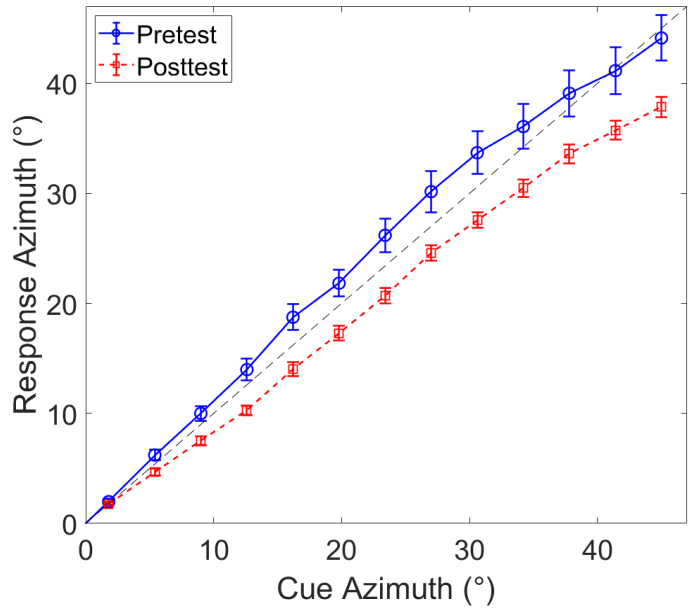

Figure 7. *Estimated response azimuths for the consistent items (i.e., the parameter Q from the regression analysis) as a function of cue azimuth, pooled across the groups. Error bars show standard errors of the mean.*

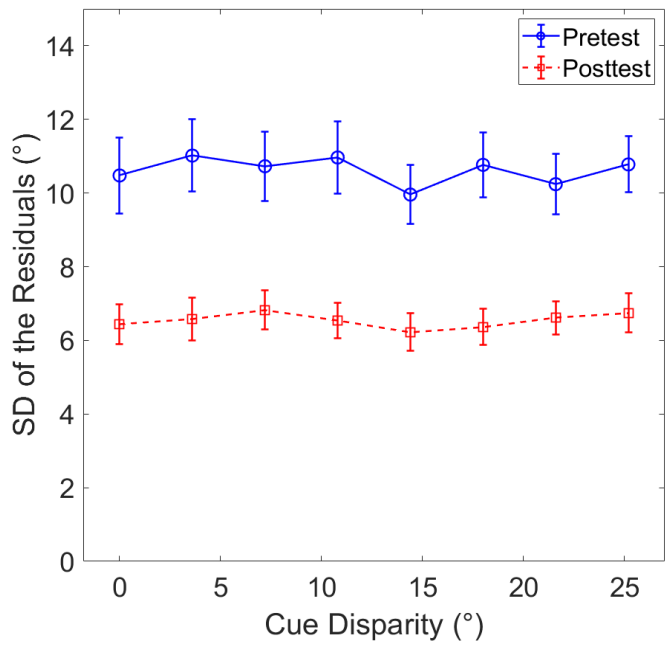

Figure 8. *Response variability (defined here as the standard deviation of the residuals of the regression analysis) as a function of cue disparity in the pre- and posttest pooled across groups. Error bars show standard errors of the mean.*